\documentclass[sigplan,screen]{acmart}
\usepackage{circledsteps}

\usepackage[linesnumbered,vlined,ruled, commentsnumbered]{algorithm2e}

\copyrightyear{2021}
\acmYear{2021}
\setcopyright{rightsretained}
\acmConference[Middleware '21]{22nd International Middleware Conference}{December 6--10, 2021}{Virtual Event, Canada}
\acmBooktitle{22nd International Middleware Conference (Middleware
'21), December 6--10, 2021, Virtual Event,
Canada}
\acmDOI{10.1145/3464298.3493388}
\acmISBN{978-1-4503-8534-3/21/12}

\begin{document}

\title{Xar-Trek: Run-time Execution Migration among FPGAs and Heterogeneous-ISA CPUs}

\author{Edson Horta, Ho-Ren Chuang, Naarayanan Rao VSathish }
\affiliation{
  \institution{Virginia Tech}
  \city{Blacksburg}
  \country{USA}
}
\email{{edsonh,horenc,naarayananrao}@vt.edu}

\author{Cesar Philippidis}
\affiliation{
  \institution{Rasec Tech}
  \city{San Jose}
   \country{USA}}
\email{cesar@rasec.tech}

\author{Antonio Barbalace}
\affiliation{
  \institution{The University of Edinburgh}
  \city{Edinburgh}
  \country{UK}}
  \email{antonio.barbalace@ed.ac.uk}

\author{Pierre Olivier}
\affiliation{
  \institution{The University of Manchester}
  \city{Manchester}
  \country{UK}
} 
 \email{pierre.olivier@manchester.ac.uk}

\author{Binoy Ravindran}
\affiliation{
  \institution{Virginia Tech}
  \city{Blacksburg}
  \country{USA}
}  
\email{binoy@vt.edu}

\renewcommand{\shortauthors}{Edson Horta, et al.}

\begin{abstract}
Datacenter servers are increasingly heterogeneous: from x86 host CPUs, to ARM or RISC-V CPUs in NICs/SSDs, to FPGAs. Previous works have demonstrated that migrating application execution at run-time across heterogeneous-ISA CPUs can yield significant performance and energy gains, with relatively little programmer effort. However, FPGAs have often been overlooked in that context: hardware acceleration using FPGAs involves statically implementing select application functions, which prohibits dynamic and transparent migration. We present Xar-Trek, a new compiler and run-time software framework that overcomes this limitation. Xar-Trek compiles an application for several CPU ISAs and select  application functions for acceleration on an FPGA, allowing execution migration between heterogeneous-ISA CPUs and FPGAs at run-time. Xar-Trek's run-time monitors server workloads and migrates application functions to an FPGA or to heterogeneous-ISA CPUs based on a scheduling policy.  We develop a heuristic policy that uses application workload profiles to make scheduling decisions. Our evaluations conducted on a system with x86-64 server CPUs, ARM64 server CPUs, and an Alveo accelerator card reveal 88\%-1\% performance gains over no-migration baselines.
\end{abstract}




\maketitle

\section{Introduction} \label{Introduction}
The increasing demand for performance, power efficiency, and reduced form factor is introducing greater architectural heterogeneity in cloud datacenter servers~\cite{ASPLOS_2017_Breaking_Boundaries,VEE_2020_H_Container}. 
Many system components, notably I/O devices such as NICs and SSDs, increasingly incorporate general-purpose CPUs whose ISAs (usually ARM or RISC-V) differ from that of host CPUs' (usually x86-64). A recent addition to this trend is reconfigurable logic~\cite{AWS_FPGA} \cite{NSDI_2018_Azure}. FPGAs have increasingly been deployed as discrete, PCIe-connected accelerator cards in NICs to boost network packet processing~\cite{Alveo_U25_SamrtNIC} \cite{ATC_2019_NICA} \cite{ATC_2019_E3} and in SSDs to enable near data processing~\cite{Samsung_SmartSSD}.
FPGAs are also increasingly co-located with other processing units or accelerators in the same chip, e.g., Xilinx Versal~\cite{Xilinx_Versal}, Intel Agilex~\cite{Intel_Agilex}.

Application development for increasingly heterogeneous hardware is non-trivial, primarily due to the fundamental differences between general-purpose, special purpose, and reconfigurable processing units. Ideally, a programmer would write an application once, and then compile and execute it among all available processing units, automatically achieving improved properties of interest such as high performance and reduced power consumption.

Recent work~\cite{ASPLOS_2017_Breaking_Boundaries, EuroSys_15_Popcorn, HDPC_19_HEXO,ISCA_2014_ISA_Diversity_Chip, ASPLOS_12_ARM_MIPS_Migration, ACM_Trans_2015_K2,OSDI_2012_COMET} has demonstrated that migrating application execution across general-purpose heterogeneous-ISA CPUs \textit{at run-time} can yield performance and energy efficiency gains. 
For example, in the datacenter multi-tenant model where several applications share the same hardware infrastructure, it may be desirable to execute software processes on power-efficient ARM CPUs during low demand, and dynamically migrate them to highly performant x86 CPUs when the workload increases. Notably, prior work has accomplished these gains with relatively little programmer involvement: applications written for a traditional shared memory programming model (e.g., POSIX, OpenMP) can be made to run on such hardware without significant custom modifications. This improved programmability is accomplished through innovations across the system software stack (i.e., operating system, compiler, run-time), which hides complexities such as ISA/ABI differences, CPUs' discrete physical memory, and cross-ISA scheduling and resource management. These works, however, have excluded FPGAs. 

In the hardware acceleration domain, application development and execution largely follows a static model: selected application functions suitable for hardware implementation are first described in high-level languages such as C or C++, or hardware description languages such as VHDL or Verilog. Using FPGA development tools, the functions are then mapped to FPGA logic resources, enabling applications to exploit CPUs and FPGAs during execution. However, once those functions have been mapped to the FPGA, they execute on the FPGA for their entire lifetime. In other words, the mapping of application to CPUs and FPGAs does not change.
This static model is highly effective for settings where the CPU/FPGA hardware is exclusively used for a single application, and the reconfigurable logic literature has intensively studied various problems in this space such as how to implement functions on FPGAs for optimal performance~\cite{FCCM_20_BQSR_Accel_Genome} \cite{HPCC_2019_Task_Sheduling_CPU_FPGA} \cite{ICPE_2019_Collaborative_Execution} \cite{FPL_2020_DSA_for_Sparse_Matrix_Mult} and how to improve FPGA programming~\cite{FPT_2019_FPGA_Overlay_RapidWright} \cite{FPL_2020_PR_for_Design_Opt}. 

In the datacenter multi-tenant model, sharing hardware resources among applications can improve overall resource utilization and   reduce hardware costs. In fact, machine acquisition costs are one of the major drivers of datacenter costs~\cite{ASPLOS_2017_Improving_DC_Efficiency}, and vendors are constantly seeking ways to reduce capital costs -- e.g., Amazon~\cite{AWS_Graviton} and Oracle~\cite{oracle-arm-cloud} recently introduced ARM-based servers in their cloud platforms to provide low-cost compute capabilities. Nonetheless, datacenter applications are increasingly computationally demanding, which are often satisfied at a better performance/density by special-purpose accelerators or FPGAs, instead of CPUs. FPGAs have several advantages over special-purpose accelerators such as customizability and lower TDP. Thus, datacenter vendors are fast incorporating FPGA-equipped servers~\cite{szefer21}, and a variety of FPGAs are now publicly available -- e.g.,  Xilinx Virtex UltraScale+ from Amazon AWS~\cite{AWS_FPGA}, Huawei Cloud~\cite{huawei}, and Alibaba Cloud~\cite{alibaba-f3}; Xilinx Kintex UltraScale from Baidu Cloud~\cite{baidu}; Xilinx Alveo Accelerators from Nimbix~\cite{nimbix}; Intel Arria 10 from Alibaba Cloud~\cite{alibaba-f1} and OVH~\cite{ovh}. 

In this paper, we explore the feasibility of dynamically provisioning FPGAs in datacenters among several tenants, in particular, as a means to alleviate \textit{dynamic} workload spikes on servers. In contrast to state-of-the-art and -practice, we consider a model in which an FPGA is not always exclusively provisioned for a client that demands and pays for acceleration. Rather, applications belonging to a multi-tenant server are allowed to use and share an attached FPGA, when the FPGA is not committed to any one client. As a result, this migration of application functions to the FPGA at run-time can help to alleviate server workload spikes and improve overall application performance. Since every application function may not have optimal execution time on CPU-FPGA hardware (e.g., for a function's compute kernel, FPGA resources may be insufficient or data transfer costs may be high), careful selection of functions for execution migration is necessary: migration should improve overall performance. 

Thus, in our model, we treat FPGAs as a special compute capability, intended for acceleration. In addition, when the FPGA is not employed for this primary use case, we propose using it to offload  functions to alleviate server workload spikes. The model therefore has the potential to not only improve overall resource utilization and performance, but also reduce hardware costs. Rather than migrate applications to another server to alleviate workload spikes, the FPGA's cost is "already paid for," as an acceleration capability for select customers. 

Such a model raises several interesting questions:

\begin{enumerate}
\item What infrastructure can enable application function migration across server CPUs and FPGAs at run-time, with relatively little programmer involvement?

\item When is dynamic execution migration to the FPGA effective? In particular, under what workload conditions is migration effective or not effective?
\item Which functions should be migrated to the FPGA to improve overall application performance, and how can these functions be selected efficiently?
\end{enumerate}

In this context, we present Xar-Trek\footnote{The name reflects the idea of ``trekking" across  \underline{X}86 and  \underline{a}rm CPUs and hardware with \underline{r}econfigurable logic.}, a compiler and run-time framework which enables execution migration of application functions from heterogeneous-ISA server host CPUs to FPGAs at run-time. We focus on performance in datacenter settings. As such, our hardware is server-grade: Intel Xeon x86 CPUs, ThunderX ARM CPUs, and an Xilinx Alveo FPGA card, interconnected using high-speed interconnects. 
%
%
Xar-Trek's compiler (Section~\ref{Compile-Time}) generates multi-ISA binaries that include FPGA implementations for a select set of application functions with very little programmer involvement. Xar-Trek's run-time (Section~\ref{Run-Time})
includes a scheduler infrastructure that monitors server workloads and migrates   functions across heterogeneous-ISA CPUs and the FPGA. 
We develop a heuristic policy that uses application workload profiles to make scheduling decisions. Our evaluation studies  (Section~\ref{Evaluation}) conducted using compute-intensive applications reveal Xar-Trek's effectiveness: performance gains in the range of 88\%-1\% over x86- and FPGA-only baselines.  

A large body of prior work has studied heterogeneous systems including CPU/FPGAs, CPU/GPUs, and heterogeneous-ISA CPUs (Section~\ref{Related Work}). However, in contrast to Xar-Trek, they do not consider heterogeneous-ISA CPUs augmented with FPGAs as a compute capability to alleviate dynamic server workload spikes in a multi-tenant setting.

Xar-Trek has several limitations (Section~\ref{sec:limitations}) including restrictions on migrate-able code, the granularity of the unit of execution migration, and  limitation to C, among others.

The paper's main contributions include:
\begin{enumerate}
\item A compiler that produces multi-ISA binaries augmented with hardware implementations of select application functions capable of migration across heterogeneous-ISA server CPUs and FPGAs at run-time.
\item A run-time system that monitors server workloads and migrates application functions across ISA-diverse CPUs and an FPGA  according to a scheduling policy.
\item A scheduling policy that aims to improve overall performance by dynamically selecting  functions for migration across ISA-diverse CPUs and an FPGA.
\end{enumerate}

\section{Background} \label{Background}

\noindent \textbf{Hardware Acceleration.}
FPGA-based hardware acceleration has been used in a variety of application domains, such as genome sequencing \cite{FCCM_20_BQSR_Accel_Genome}, clustering algorithms \cite{FPGA_2020_BiS-KM}, and databases \cite{FPL_2018_Accel_Database_Survey}. The development of applications that can take advantage of hardware accelerators often depends on the expertise of hardware designers regarding complex tools, although there is significant automation in generating configuration files (bitstreams) and integrating them with host applications  \cite{FPL_20_Vitis_for_HPC}. 
The traditional approach to hardware acceleration always targets the FPGA to execute accelerated functions. Software versions of select application functions are converted into hardware kernels with the aid of EDA tools such as Xilinx's Vitis~\cite{Vitis_Doc}. Thereafter, the host application is instrumented to transfer the input data to the FPGA, call the function in hardware, wait for the execution to complete, and transfer the results back to the host CPU. When the hardware version of the selected function is slower than its software version, the hardware kernel is discarded. 

\noindent \textbf{Heterogeneous-ISA Platforms.}
Traditionally, compiled applications cannot migrate at run-time across heterogeneous-ISA CPUs because of the ISA difference. Recent research (e.g., Popcorn Linux~\cite{ASPLOS_2017_Breaking_Boundaries, EuroSys_15_Popcorn, 
HDPC_19_HEXO},
Venkat and Tullsen~\cite{ISCA_2014_ISA_Diversity_Chip, ASPLOS_12_ARM_MIPS_Migration}, K2~\cite{ACM_Trans_2015_K2}, Comet~\cite{OSDI_2012_COMET}) overcome this through multi-ISA compilation and cross-ISA program state transformation at run-time. After source code is lowered to an intermediate representation, the compilation process inserts ``migration points" where program has equivalent memory state across ISAs~\cite{ACM_Tras_1994_Equivalence_Point}  and therefore execution migration across ISAs is possible. At run-time, the migration points call-back into a run-time library, which, according to a scheduling policy, makes migration decisions~\cite{ASPLOS_2017_Breaking_Boundaries}. The compiler generates ISA-specific machine code for each ISA-different CPU, and aligns all symbols (i.e., globals, statics, functions) at the same virtual address across all ISAs, for uniform meaning of addresses. Metadata necessary for transforming the program state at run-time (e.g., live variables at call sites) are also generated (i.e., by a liveness pass). At run-time, when a migration decision is made at a migration point, the run-time library transforms the program's dynamic state that is ISA-specific (e.g., stack, registers) from the source ISA format to the destination ISA format, leveraging the metadata. 

In heterogeneous-ISA hardware with no shared memory between ISA-different CPUs, prior work~\cite{ASPLOS_2017_Breaking_Boundaries, ACM_Trans_2015_K2} implements distributed shared memory (DSM)~\cite{CMU_2018_DSM} as a first-class OS abstraction. DSM provides sequentially-consistent memory state across (ISA-different) CPUs. Thus, once the run-time library transforms the program state, thread or process execution can resume on the destination ISA, observing the same program order as that on the source ISA.

Since the Popcorn Linux~\cite{ASPLOS_2017_Breaking_Boundaries} infrastructure is publicly available, our work builds on top of it with a compiler and run-time for x86 CPUs, ARM CPUs, and FPGAs.

\section{The Xar-Trek Compiler and Run-Time Framework} \label{Framework}

At a high-level, Xar-Trek's compiler framework works as follows. An application is first profiled to determine the functions that can be executed on the three architectures. For each function that can be implemented in hardware, a subsequent step instruments  the code to insert calls to Xar-Trek's scheduler (which makes scheduling decisions) and to an FPGA configuration function (which pre-configures the FPGA for the function). The Popcorn Linux compiler~\cite{ASPLOS_2017_Breaking_Boundaries} is then invoked on this instrumented code to produce multi-ISA binaries that can execute on x86-64 and ARM64 CPUs. 

In the next step, the functions that can be implemented in FPGA, as identified in the profiling step, are mapped to the FPGA using the Xilinx compiler, generating hardware object files, which are then used to generate FPGA configuration files, followed by generating hardware kernels, which are downloaded to the FPGA. 

In the compiler's final step, applications are executed in different migration scenarios and x86 CPU load thresholds at which execution migration to ARM CPU or FPGA is likely effective, is determined. Xar-Trek's run-time system dynamically refines these thresholds based on run-time behaviors. 

Xar-Trek's run-time system contains a userspace scheduler client and a scheduler server, which are integrated with application functions during the compiler's instrumentation step. The client executes after functions return from their execution and uses the observed run-time execution times to refine the statically determined x86 CPU load thresholds. The scheduler server is invoked before function calls and  makes policy decisions on where to execute functions.

We now explain the framework in detail. 
 
\subsection{The Compiler Framework} \label{Compile-Time}

\begin{figure*}[!ht]
\centering
\includegraphics[width=5in]{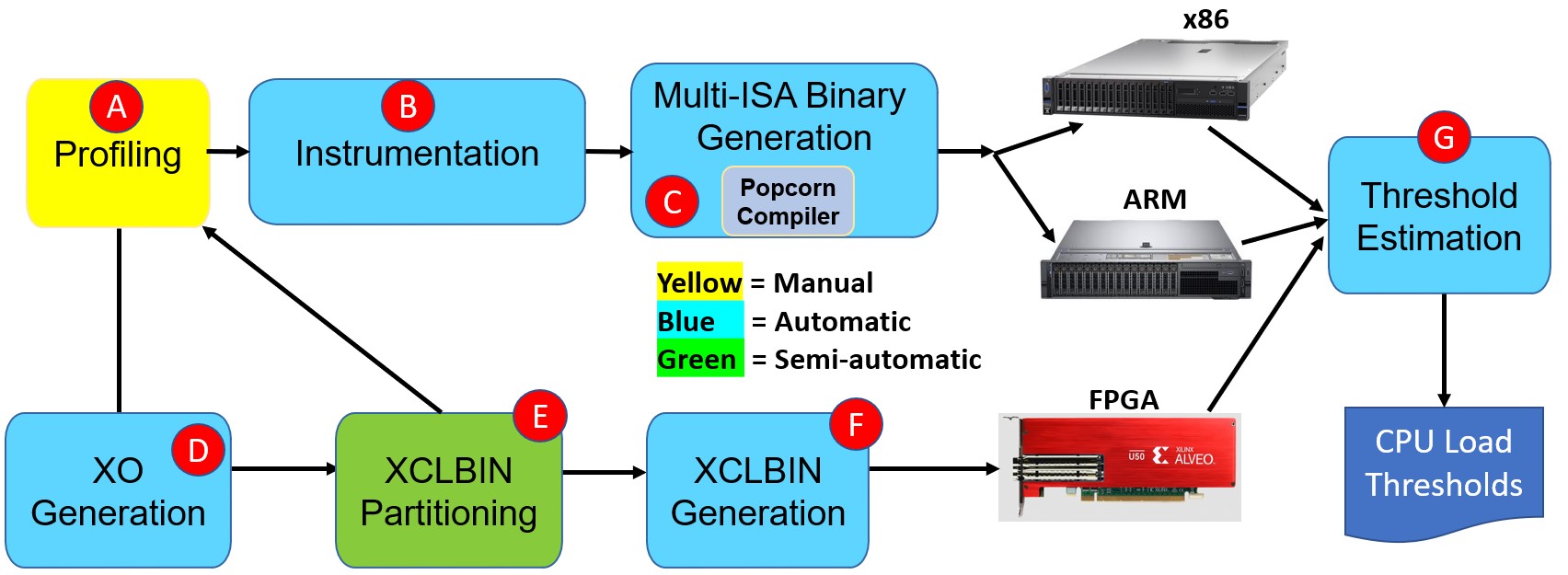}
\caption{The Xar-Trek compiler framework.}
\label{Meth_Compile}
\end{figure*}

Figure~\ref{Meth_Compile} illustrates Xar-Trek's compiler framework. Each step of the framework, \Circled{A}-\Circled{G}, involves a separate program. All steps are unique to Xar-Trek, except the \textit{Multi-ISA Binary Generation} step  \Circled{C}, which is leveraged from Popcorn Linux. 


The first step, \textit{Profiling} (\Circled{A} in Figure~\ref{Meth_Compile}), is a manual step performed by an application designer to define the function(s) that can be executed on any of the three target architectures. Profiling tools such as gprof \cite{SIGPLAN_2004_GPROF} and valgrind \cite{PLDI_07_VALGRIND} can be used to assist with this task. This manual step's outcome is a text file which describes: 1) the hardware platform; 2) the applications; and 3) the selected functions of each application. This task is very important due to the fact that the switch from software to hardware is done at function boundaries.
We made this design decision as current Xilinx tools only support self-contained functions and do not allow converting only parts of a function to hardware.\footnote{Extending Xilinx tools to support this functionality is outside the paper’s scope. In some situations, functions can be outlined to overcome this limitation. Note that the paper’s contribution is orthogonal to this: Xar-Trek’s compiler/run-time infrastructure and scheduling policy are independent of the granularity of the unit of execution migration.} 

The next two steps, \textit{Instrumentation} \Circled{B} and \textit{Multi-ISA Binary Generation} \Circled{C},  are fully automated and generate executable files for both x86-64 and ARM64 CPUs. Xar-Trek's scheduler is implemented using a client/server architecture. For each application function selected for implementation in hardware, the instrumentation step inserts calls for the scheduler client in the application to communicate with the scheduler server, using sockets and signals (see Section~\ref{Run-Time}). These calls are placed at the beginning and at the end of the application's main function. 
In addition, at the main function's start, the tool inserts a call to a function that configures the FPGA and prepares it to run each selected functions when needed. This technique allows the hardware kernel to be called without having to wait for its initialization. As we show in Section~\ref{Evaluation}, this is important for applications that use hardware kernels more than once. The instrumentation step also replaces the original call of the selected functions with calls to different targets (x86, ARM, and FPGA) according to a flag set by the scheduler client. 

The instrumented application is the input to the \textit{Multi-ISA Binary Generation} step \Circled{C}, which invokes the Popcorn compiler to generate binaries for x86-64 and ARM64 CPUs. 

The file produced by the profiling step is the input to the \textit{Xilinx Object  Generation} step \Circled{D}. This step automatically moves the selected functions for hardware implementation to new files and invokes the Xilinx Vitis compiler to map them to the FPGA, generating one Xilinx Object (XO) file for each function. Vitis automatically performs this mapping. Although HLS pragmas can be manually inserted in the source code, our work primarily targets software developers who may not have experience with high-level synthesis. Thus, we rely on off-the-shelf tools to optimize hardware kernels. Our premise is that users (and datacenter operators) want to take advantage of FPGA resources (when they are available) for their applications, and in doing so, trust optimizations provided by FPGA development tools such as Vitis. At the same time, our toolchain allows the use of any optimized hardware libraries already available from FPGA vendors.  

The next step, \textit{XCLBIN Partitioning}  \Circled{E}, gathers information about the FPGA resource utilization from the XO files and the area available in the hardware platform to estimate how many functions can be grouped in one configuration file. The hardware platform contains all the static hardware modules inside the FPGA, comprising of the host CPU interface, reconfiguration control, memory controllers, and reserved areas to receive the hardware kernels. The hardware platform and all the selected functions, present in the XO files, are integrated into one binary file, called XCLBIN, and is downloaded into the FPGA. In the event that more than one XCLBIN is needed to host all the selected functions, the tool automatically assigns them to multiple XCLBIN files. This automatic partitioning can also be manually performed by assigning each function to a specific XCLBIN file, allowing the designer to iteratively define the higher priority functions that will be assembled in the same XCLBIN file. 

Subsequently, the \textit{XCLBIN Generation} step \Circled{F} uses the assignment of functions to XCLBIN(s) to call the Vitis compiler to implement the hardware kernels. The XCLBIN(s) are then downloaded to the FPGA platform.

The \textit{Threshold Estimation} step \Circled{G}  enables Xar-Trek's scheduler to efficiently utilize the hardware kernels. First, the total execution time of each application, in isolation, is measured in two migration scenarios: 1) x86-to-ARM and 2) x86-to-FPGA. 
By measuring the total execution time with migration included, we ensure that all communication overhead inherent to hardware acceleration is accounted for when the scheduler decides where to execute each function at run-time. 
Subsequently, the estimation tool executes each application on the x86 CPU while increasing the CPU load, until the application's execution time exceeds the previously recorded execution times for the two migration scenarios, x86-to-ARM and x86-to-FPGA. (The CPU load is increased by executing, in parallel, new instances of the same application.)  Our rationale is that by determining the CPU load at which an application's x86 execution time exceeds its x86-to-ARM execution time and x86-to-FPGA execution time, we can reasonably estimate the load at which migration is likely effective.
The tool records these CPU loads as ``threshold values" to trigger execution migration to ARM and FPGA, respectively. The tool outputs a table that describes, for each application, 1) the application name, 2) the hardware kernel of the application's function, 3) the FPGA threshold, and 4) the ARM threshold.

The execution of an application's selected functions on the ARM CPU or on the FPGA board incurs a communication overhead when compared to executing the entire application only on the x86 CPU. In order to migrate the function to the ARM CPU, it is necessary to first transform the application data and the program state from the source ISA/ABI format to the destination ISA/ABI format\footnote{This state transformation is done by the Popcorn Linux run-time.} and then transfer the data and the state using a communication channel such as the Ethernet interface (using its driver). Since this channel is shared among all the running processes, it is non-trivial to estimate the communication overhead. In the FPGA case, the communication overhead involves preparing the application data to be sent to the device,\footnote{In the FPGA case, state transformation is not necessary as migration from software to hardware occurs only at function boundaries, and the function's hardware implementation operates on self-contained, in-memory data.} and the time to send/receive this data to the board, usually through the PCIe interface. As in the ARM case, this interface is also shared and estimating the communication cost is a challenge. The approach adopted by Xar-Trek is to measure the execution time \textit{in locus}. By doing so, we ensure that all communication overhead inherent to function migration to ARM/FPGA, even when the data-transfer interfaces are different (Ethernet or PCIe), is accounted for when the scheduler decides where to execute each function at run-time. 

Note the Xar-Trek compiler's clear delta over the Popcorn Linux compiler~\cite{ASPLOS_2017_Breaking_Boundaries}: all steps \Circled{A}-\Circled{G} are unique to the Xar-Trek compiler, except the \textit{Multi-ISA Binary Generation} step  \Circled{C}, which is leveraged from Popcorn Linux. 

\subsection{Run-Time System} \label{Run-Time}

\begin{figure}[!ht]
\centering
\includegraphics[width=0.8\columnwidth]{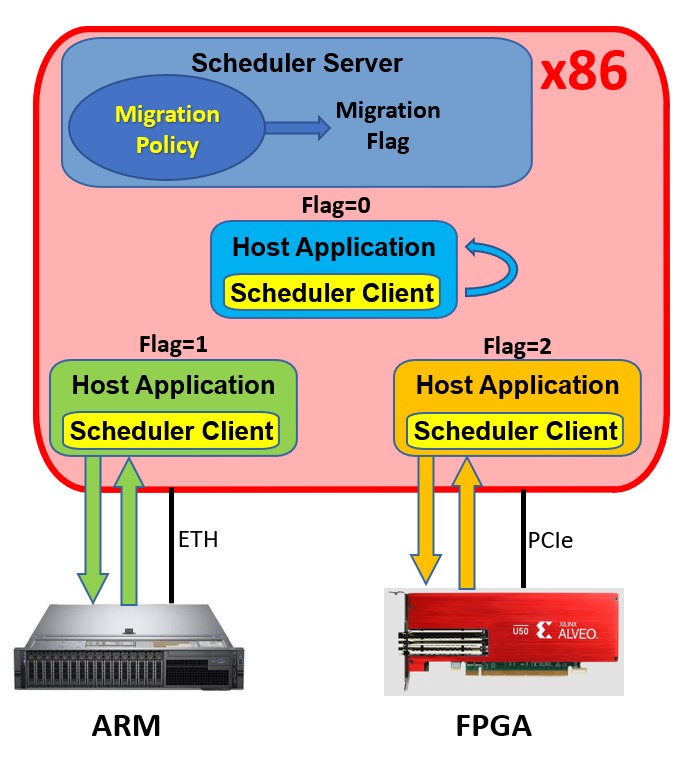}
\caption{Xar-Trek's run-time system for a heterogeneous-ISA platform with x86 and ARM CPUs and an FPGA. Flag equals target ID.}
\label{meth-run-time}
\end{figure}

Xar-Trek's run-time system's main component is a user-space scheduler that dynamically decides where each application function should run. This run-time system is integrated with Popcorn Linux's run-time library. Figure~\ref{meth-run-time} illustrates a high-level view of the system. 

The scheduler is implemented using a client/server model. An instance of the scheduler client is integrated with each application binary (x86, ARM). The scheduler client implements a procedure (see Section~\ref{subsec:threshold-update}) to dynamically update the threshold table generated during compilation. The scheduler server, which encapsulates the scheduling policy (see Section~\ref{subsec:sched-policy}), runs on the x86 host. The clients and the server communicate with each other to decide when and where to migrate applications' functions. 

Applications are initially launched on x86 host CPUs, one process per CPU. A scheduler client instance is also integrated with the base Popcorn run-time. This scheduler instance monitors a migration flag to decide a function's migration target: 0) x86 (do not migrate); 1) ARM (migration to ARM CPU, i.e., software migration); and 2) FPGA (hardware migration). Software migration is already implemented by Popcorn's run-time. Hardware migration employs OpenCL APIs in the Xilinx Runtime Library (XRT) \cite{Xilinx_XRT} to: 1) configure the hardware accelerator card; 2) manage data movement between the host and the accelerator card; and 3) orchestrate function execution.



\subsection{Dynamic Threshold Update}
\label{subsec:threshold-update}

The scheduler client instance implements Algorithm~\ref{alg:threshold} to dynamically update the threshold table that was generated during compilation (Section~\ref{Compile-Time}).

The algorithm starts by recording the data every time a function returns from its call (lines 1-2). After that, if the application is running on the x86 CPU, it compares the execution time against the FPGA execution time and the x86 CPU load against the FPGA threshold. As a result of this comparison, the FPGA threshold is updated with the x86 CPU load (lines 4-5). If the x86 execution time is greater than the ARM execution time, the algorithm checks the x86 CPU load to verify that it is smaller than the ARM threshold and, if affirmative, updates the ARM threshold (lines 7-8). If none of these conditions are met, the algorithm only updates the x86 execution time (line 10).

\begin{algorithm}
\SetAlgoLined
  Record application execution time\;
  Record CPU Load\;
    \eIf{ Application executed on x86}{
    \eIf{(x86\textsubscript{exec} $>$ FPGA\textsubscript{exec})  {\bf and} 
  (x86\textsubscript{LOAD} $<$ FPGA\textsubscript{THR})}{
   FPGA\textsubscript{THR} = x86\textsubscript{LOAD} \;
  }
  {
   \eIf{(x86\textsubscript{exec} $>$ ARM\textsubscript{exec})  {\bf and} 
  (x86\textsubscript{LOAD} $<$ ARM\textsubscript{THR})}{
  ARM\textsubscript{THR} = x86\textsubscript{LOAD} \;
  }{  Record x86\textsubscript{exec}\;
   }
  }

   }
   {\eIf{ Application executed on ARM}{
       \If{(ARM\textsubscript{exec} $>$ x86\textsubscript{exec}) }{
   Increase ARM\textsubscript{THR} \;
  }
  
  {
   
  }
  }{\If{ Application executed on FPGA}{
       \If{(FPGA\textsubscript{exec} $>$ x86\textsubscript{exec}) }{
   Increase FPGA\textsubscript{THR} \;
  }
  }}
  }
  
 \caption{Xar-Trek's dynamic threshold update procedure.}
 \label{alg:threshold}
\end{algorithm}

If the application is running on the ARM CPU, the scheduler checks if the execution time on this CPU is greater than that on the x86 CPU recorded before, and, if affirmative, increases the ARM threshold (lines 14-17). 

The last case is for when the application is running on the FPGA. The scheduler increases the FPGA threshold when the FPGA execution time is greater than the x86 execution time (lines 19-23).

By measuring the execution time immediately before the application terminates, the scheduler is able to use real-time data to update the threshold table. This increases the accuracy of the information that the scheduler  can use in its policy decisions about where to execute an application function for its next invocation.

\subsection{Scheduling Policy}
\label{subsec:sched-policy}

Xar-Trek's scheduler uses a heuristic policy, which is implemented in the scheduler server. The scheduling policy's pseudo-code is shown in Algorithm~\ref{alg:scheduler}.

\begin{algorithm}
\SetAlgoLined
  Query Available HW Kernels in the FPGA\;
  Set up socket communication\;
  Start timer to read x86\textsubscript{LOAD}\;
 \While{1}{
  Wait for client scheduler request\;
  \If{ Application invokes client scheduler}{
  Accept Connection from Application (Client)\;
  Read Threshold Table (ARM\textsubscript{THR}; FPGA\textsubscript{THR})\;
  \If{(x86\textsubscript{LOAD} $<=$ ARM\textsubscript{THR})  {\bf and} 
  (x86\textsubscript{LOAD} $>$ FPGA\textsubscript{THR}) {\bf and} 
  (No HW Kernel) }{
   Continue on x86\;
   Reconfigure the FPGA\;
   Query Available HW Kernels in the FPGA\;
  }
   
  \If{(x86\textsubscript{LOAD} $>$ ARM\textsubscript{THR}) {\bf and} (x86\textsubscript{LOAD} $>$ FPGA\textsubscript{THR}) {\bf and} (No HW Kernel) }{
   Migrate to ARM\;
   Reconfigure the FPGA\;
   Query Available HW Kernels in the FPGA\;
   }
   
  \If{(x86\textsubscript{LOAD} $<=$ ARM\textsubscript{THR}) {\bf and} (x86\textsubscript{LOAD} $<=$ FPGA\textsubscript{THR}) }{
   Continue on x86\;
   }
   
  \If{(x86\textsubscript{LOAD} $>$ ARM\textsubscript{THR}) {\bf and} (x86\textsubscript{LOAD} $<=$ FPGA\textsubscript{THR}) }{
   Migrate to ARM\;
   }
   
  \If{(x86\textsubscript{LOAD} $>$ FPGA\textsubscript{THR}) {\bf and} (HW Kernel Available) }{
   {
   \eIf{(FPGA\textsubscript{THR} $<$ ARM\textsubscript{THR})  }{
   Migrate to FPGA\;
   }{
   Migrate to ARM\;
   }
   
   }
   }
   }
   
 }
 \caption{Xar-Trek's scheduling policy.}
 \label{alg:scheduler}
\end{algorithm}

During initialization (lines 1-3), information about hardware kernels embedded in the XCLBIN file is collected, the socket connection is stabilized, and the timer used to obtain the x86 CPU load is started. This data is used in conjunction with the threshold table generated by the estimation tool to determine targets for select functions. 

The main loop (lines 4-33) waits for a request from each application and, when accepted, reads the threshold table (lines 7-8). If the hardware kernel is not present on the FPGA and the x86 load is greater than the FPGA threshold, the scheduler decides to execute the function in x86 (lines 9-13) or in ARM (lines 14-18), depending on the ARM threshold value. In both cases, while it is executing on one of the targets, the scheduler reconfigures the FPGA and updates the information about the available hardware kernels in the FPGA (lines 11-12 and 16-17). Until the reconfiguration is complete, the function remains on the x86 CPU or may migrate to the ARM CPU. This strategy allows us to hide the transfer and reconfiguration latencies. 

If the x86 load is less than the ARM/FPGA thresholds, the function is executed on x86 (lines 19-21). When the x86 load exceeds only the ARM threshold, the scheduler executes the function on the ARM CPU (lines 22-24). These last two options happens when the x86 load exceeds the FPGA threshold and the hardware kernel is present in the FPGA (lines 25-31). If the FPGA threshold is less than the ARM threshold, the function will be executed in the FPGA (lines 26-27); otherwise, the function will be executed in ARM (lines 29-30). This is due to the fact that the smaller threshold implies that the target has a smaller execution time for that function.

\section{Evaluation} \label{Evaluation}

Our evaluation hardware is the same as shown in Figure~\ref{meth-run-time}, and consists of a Dell 7920 server (Xeon Bronze 3104 CPU, 1.7GHz, 6 cores, 64GB), a Cavium ThunderX server (ARM CPU, 2GHz, 96 cores, 128GB), and a Xilinx Alveo U50 card~\cite{Alveo_U50}. Both servers run Ubuntu 18.04.5 LTS with Popcorn Linux kernel 4.4.137. The interconnects are Ethernet for the servers ($1$Gbps) and PCIe ($32$GB/s) for the FPGA.

We focused on AI/vision and HPC workloads as representative of  compute-intensive datacenter applications and used the following Rosetta benchmarks~\cite{FPGA_2018_Rosetta}: face detection (used in image processing \cite{TNNLS_2019_Object_Detection}) and digit recognition (used  in machine learning~\cite{ASSSC_2018_Handwritten_Digit_Recognition}). Face detection targets image sizes of 320x240 (original implementation; referred to as `FaceDet320') and 640x480 (`FaceDet640'). We updated this benchmark to support multiple images of size 320x240  to evaluate throughput. The digit recognition benchmark was used with both 500 (`Digit500') and 2000 (`Digit2000') tests.

We also used HPC workloads from NAS Parallel Benchmark (NPB)~\cite{SuperComp_1991_NPB} suite's CG-A application as representative of a class of applications that are significantly slower on the FPGA than on x86. This benchmark is also the only one in our evaluation that runs faster on the ARM CPU than FPGA.

\begin{table}[!ht]
\renewcommand{\arraystretch}{1.7}
\caption{Benchmark execution times (milliseconds).}
\label{Exec_Time_App}
\centering
\begin{tabular}{c | c | c | c}
\hline
Benchmark & Vanilla Linux & Xar-Trek & Xar-Trek \\
 & \footnotesize{(x86 only)} & \footnotesize{(x86/FPGA)} & \footnotesize{(x86/ARM)}\\
\hline
CG-A                    & 2182  & 10597  & 8406\\
\hline
FaceDet320      & 175   & 332   & 642\\
\hline
FaceDet640      & 885   & 832   & 2991\\
\hline
Digit500    & 883   & 470   & 2281 \\
\hline
Digit2000  & 3521  & 1229  & 8963\\
\hline
\end{tabular}
\end{table}

Xar-Trek was used to compile each of the following five benchmarks, while exploiting Vitis 2020.2 to generate the XCLBIN hardware kernels: CG-A; face detection (320x240); face detection (640x480); digit recognition (500 tests); and digit recognition (2000 tests). The face detection kernel with an image size of 320x240 was also used in the updated version that allows the user to choose the number of images processed.

Table~\ref{Exec_Time_App} shows each benchmark's execution time when the selected function is executed on x86 without using Xar-Trek (``Vanilla Linux"), along with the execution time when the function migrates to FPGA or to ARM, using Xar-Trek.

\begin{table}[!h]
\renewcommand{\arraystretch}{1.7}
\caption{Xar-Trek's threshold estimation.}
\label{Threshold_Prediction}
\centering
\begin{tabular}{c | c | c | c }
\hline
Benchmark  &  HW Kernel  &  FPGA\textsubscript{THR}  &  ARM\textsubscript{THR}\\
\hline
CG\_A  &  KNL\_HW\_CG\_A  &  31  &  25 \\
\hline
FaceDet320  &  KNL\_HW\_FD320  &  16  &  31 \\
\hline
FaceDet640  &  KNL\_HW\_FD640  &  0  &  23 \\
\hline
Digit500  &  KNL\_HW\_DR500  &  0  &  18 \\
\hline
Digit2000  &  KNL\_HW\_DR200 &  0  &  17 \\
\hline
\end{tabular}
\end{table}

The original face detection benchmark stores the image in an executable file. The software function makes use of internal buffers to store this image. As the image's size increases, the hardware implementation outperforms x86 (Table~\ref{Exec_Time_App}'s second and third rows), since it is using FPGA's internal memories.
Table~\ref{Threshold_Prediction} shows the data generated by Xar-Trek's threshold estimation tool, using the execution times when the function does not migrate and when it migrates to FPGA or to ARM.

\begin{table}[!h]
\renewcommand{\arraystretch}{1.7}
\caption{CPU load definition.}
\label{CPU-LOAD}
\centering
\begin{tabular}{c | l  }
CPU Load  &  Range of number of processes\\
\hline
Low  &  \#processes $<$ \#x86 cores  \\
\hline
Medium & \#processes $>$ \#x86 cores \\
       & \#processes $<$ (\#x86 cores + \#ARM cores) \\
\hline
High & \#processes $>$ (\#x86 cores + \#ARM cores)  \\
\hline
\end{tabular}
\end{table}

Since the ratio of the number of application processes to the number of available cores is a reasonable measure of CPU load (for compute-intensive applications), we defined three CPU loads using this metric: low, medium, and high, shown in Table~\ref{CPU-LOAD}. Note that the total number of cores available is 102 (6 x86 cores and 96 ARM cores). We used the five benchmarks, where the selected function is called once per run, inside each application.  

\subsection{Performance: Average Execution Time}

Our first goal was to understand Xar-Trek's effectiveness in improving average execution time of an application set during low loads, i.e., when the number of processes does not exceed the number of x86 cores, and additional compute cycles are usually not needed. During such load situations, can Xar-Trek perform as well as an x86-only solution? 

To avoid selection bias, we randomly selected (using an  uniform distribution) a set of 1, 2, 3, 4, and 5 applications from our five benchmarks, and measured the set's average execution time for Xar-Trek and compared it against two baselines: \textit{Vanilla Linux/x86} (always executed on x86), and  \textit{FPGA} (always executed on FPGA). 
For this evaluation, the CPU load is equal to the number of applications running in the system. 

\begin{figure}[!h]
\centering
\includegraphics[width=3.3in]{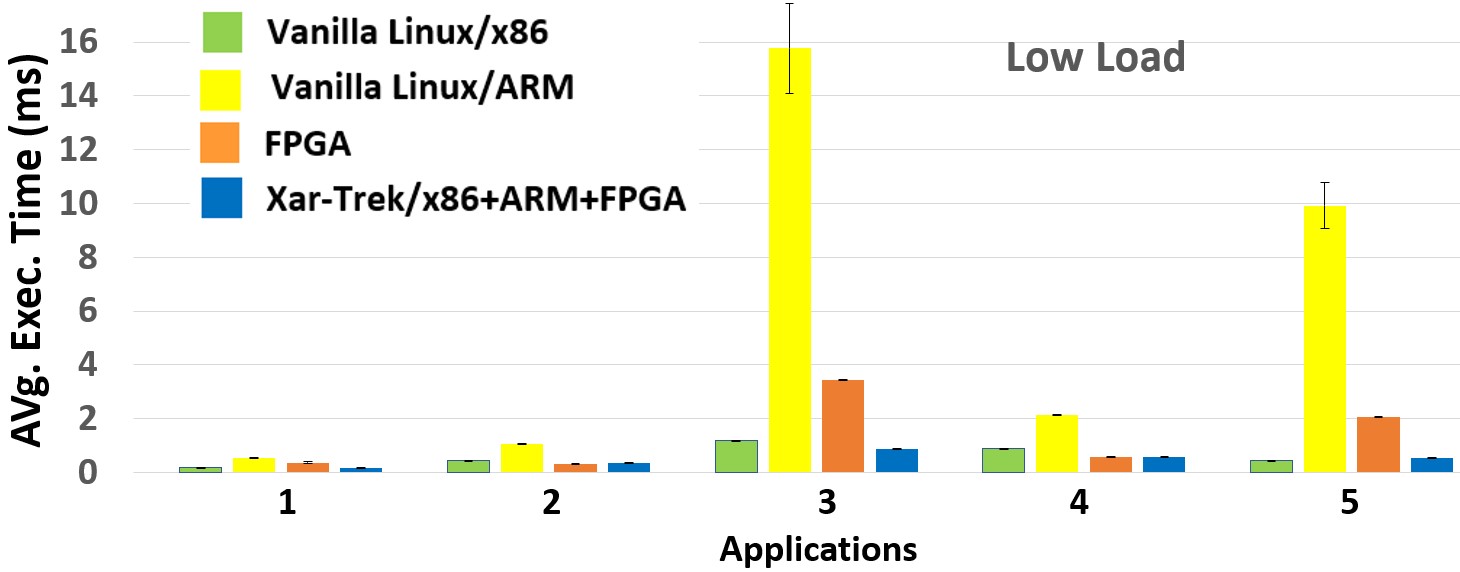}
\caption{Average execution time of a randomized application set with less processes than the \#x86 cores in the hardware. Lower is faster.}
\label{Low_Load}
\end{figure}

\begin{figure}[!h]
\centering
\includegraphics[width=3.3in]{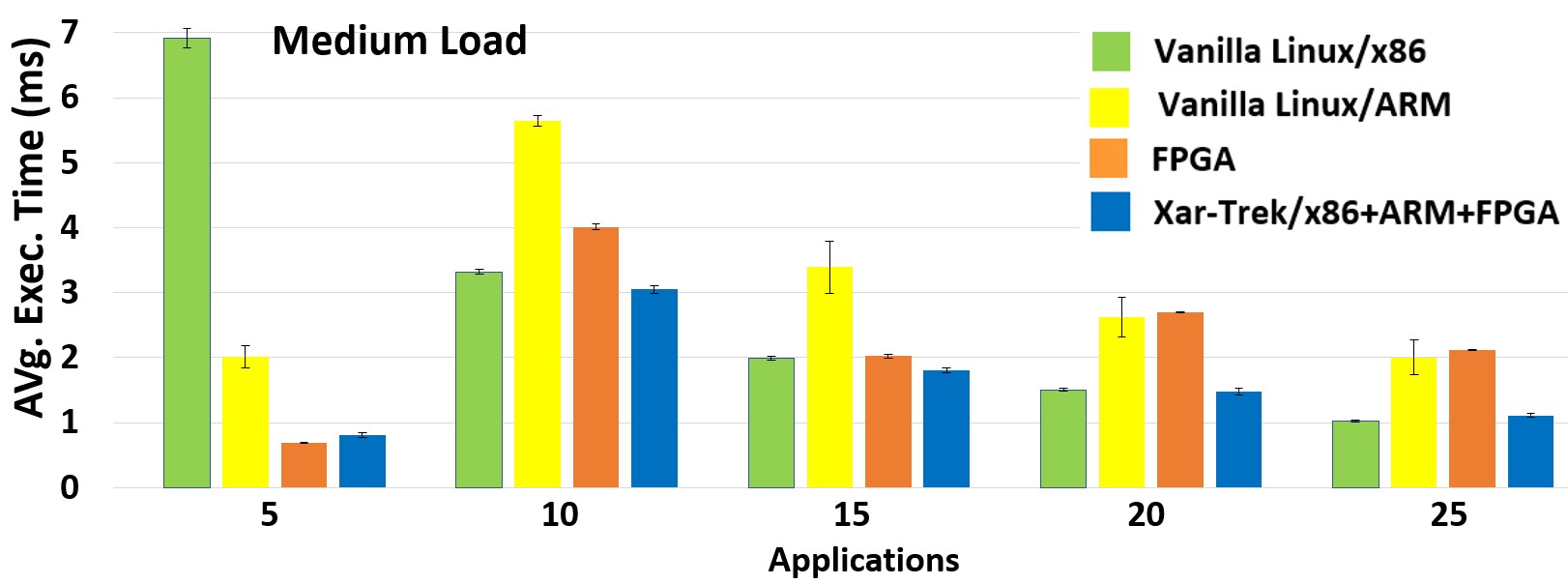}
\caption{Average execution time of a randomized application set with 60 processes (more than \#x86 cores, but less than total \#cores). Lower is faster.}
\label{Medium_Load}
\end{figure}

\begin{figure}[!h]
\centering
\includegraphics[width=3.3in]{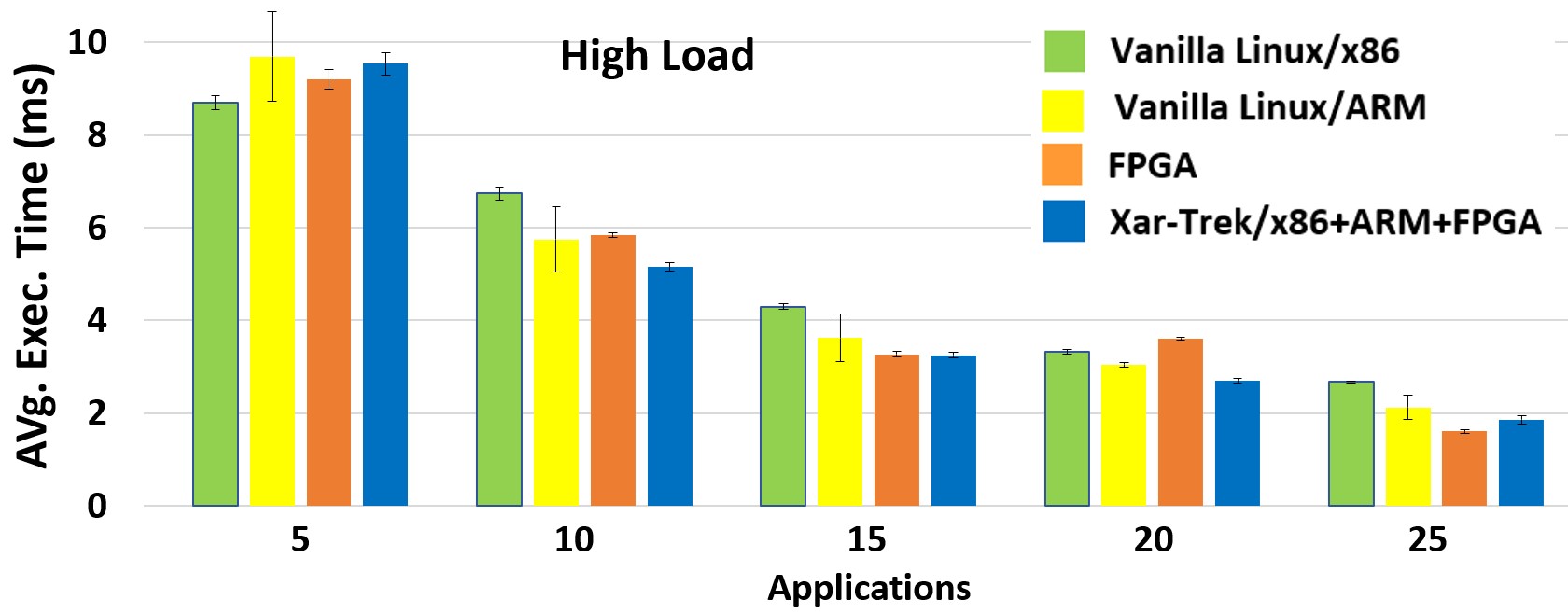}
\caption{Average execution time of a randomized application set with 120 processes (more than total number of cores). Lower is faster.}
\label{High_Load}
\end{figure}

Figure~\ref{Low_Load} shows the results. In all figures (Figures~\ref{Low_Load}-\ref{face-detect-throughput}), each data point is the average of $10$ runs and the standard deviation is nearly zero (shown as small vertical lines at the top of each bar).
The results indicate that Xar-Trek is always superior (except in two cases), with Vanilla Linux/x86 performing the same or a close second, revealing that Xar-Trek does \textit{not} migrate in most cases, which is effective. The Vanilla Linux/ARM bar represents the case when the application is executed only on the ARM server, which is always slower than the other cases. Xar-Trek's largest and smallest gains, compared to FPGA, are  75\% and 50\%. 
In the 5-application case, Vanilla Linux is the best, by 21\%. 
This is because Xar-Trek is slower than Vanilla Linux for CG-A or Face Detection (320x240), which are present in this case.

The results also reveal that when applications always run on the FPGA, performance degrades significantly when there is one application that is slower on the FPGA. In other words, the traditional approach of always using FPGA is not \textit{always} effective in a shared tenant setting during low loads.

Our second goal was to understand Xar-Trek's effectiveness when the number of processes exceeds the number of x86 cores, but not the total number of cores (medium load), and when the number of processes exceeds the total number of cores (high load). In these cases, additional compute cycles are needed -- can Xar-Trek migrate and yield gains?

To avoid selection bias, we again randomly grouped applications in sets of 5, 10, 15, 20, and 25 from the five benchmarks. CPU load was generated by running simultaneously the NPB MG-B application $n$ times while executing each set.

Figures \ref{Medium_Load}-\ref{High_Load} show the results.  
Xar-Trek almost always outperforms Vanilla Linux/x86 in medium (largest and smallest gains: 88\% and 1\%) and high loads (largest and smallest gains: 31\% and 19\%). 
These results reveal Xar-Trek's effectiveness in dynamically selecting the right function and the right target in order to best leverage the heterogeneous resources. 


\subsection{Performance: Throughput}

Our third goal was to understand whether Xar-Trek can improve the throughput of applications that call a selected function multiple times, an important metric of interest in datacenters. To measure this, we modified the face detection application by allowing the user to choose the number of images processed. This choice impacts the number of times a selected function is called. In the original benchmark, the image was embedded in the binary file. The modified version reads each file before processing it. The images were obtained from the WIDER dataset~\cite{Wider_Images} and converted to PGM format. Since the results from low, medium, and high loads (Figures~\ref{Low_Load}-\ref{High_Load}) revealed Vanilla Linux/ARM's inferior performance, we excluded it in this evaluation.

\begin{figure}[!ht]
\centering
\includegraphics[width=3.3in]{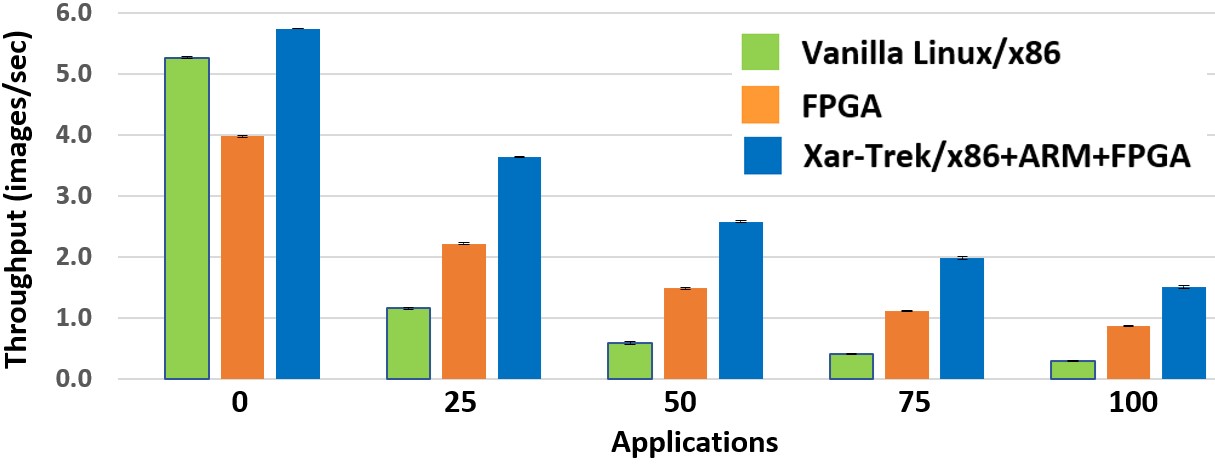}
\caption{Throughput of face detection. Higher is better.} 
\label{face-detect-throughput}
\end{figure}

The CPU load was generated using the same approach of the previous cases, with \textit{n} equalling 0, 25, 50, 75, and 100 processes. Face detection was set to run on 1000 images for 60 seconds, after which the application was terminated, and the number of processed images was counted. We used the same scheduling policy and baselines as before.

The results are shown in Figure~\ref{face-detect-throughput}, which reveal Xar-Trek's throughput improvements when the CPU load exceeds 25 processes (average gain is $\approx$4x). 
This is because the FPGA threshold in Table~\ref{Threshold_Prediction}'s second row is 16. Thus, when the system load exceeds 25 processes, Xar-Trek migrates the function to the FPGA.

The results also demonstrate that Xar-Trek is even faster than the always-FPGA baseline. This is a direct consequence of configuring the FPGA at the very beginning of the application, as discussed in Section~\ref{Framework}.

\subsection{Performance: Periodic Workload}

The experiments reported so far (i.e., Figures~\ref{Low_Load}-\ref{face-detect-throughput}) evaluated Xar-Trek's performance for a fixed workload. Although fixed workloads allow us to understand Xar-Trek's first-order performance, they are not representative of datacenter workload patterns which have time-varying behaviors~\cite{ASPLOS_2017_Improving_DC_Efficiency}, i.e., job arrivals vary over time. To understand Xar-Trek's effectiveness for such dynamic workloads, in particular, how the scheduler can make effective decisions as the number of processes quickly climb from a medium load to a high load and then decay to a medium load, inspired by previous works~\cite{ASPLOS_2017_Breaking_Boundaries}, we conducted experiments with a \textit{periodic} workload that generated 20 processes (medium) to 160 processes (high) over time: during an experimental time frame of 43 minutes, thirty sets of 20 applications from our five benchmarks 
were launched with an interval of 30 seconds per set, forming a wave-like load pattern.


\begin{figure}[!ht]
\centering
\includegraphics[width=3.0in]{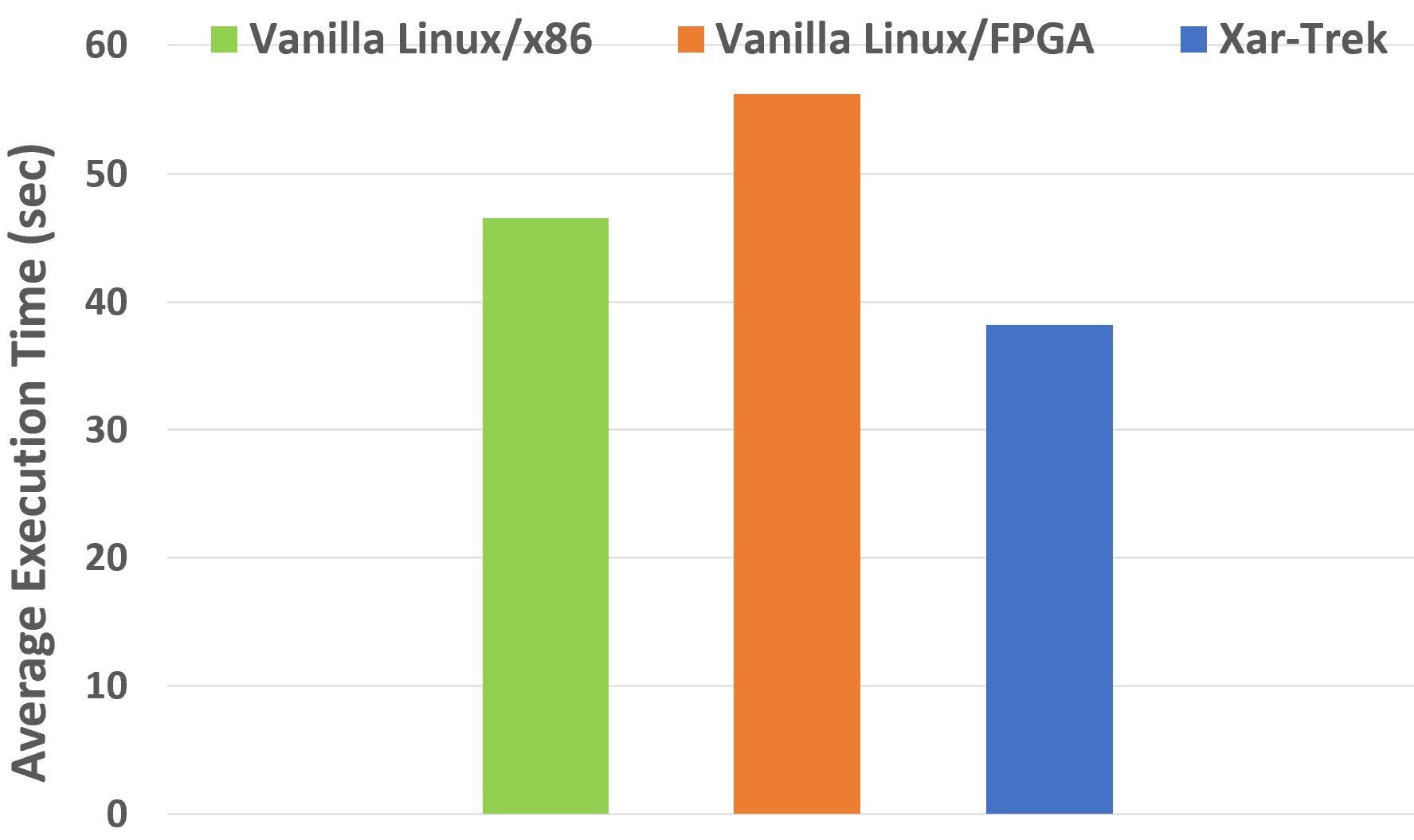}
\caption{Average execution time of thirty waves of 20 applications,  forming a periodic workload. Lower is better.} 
\label{continuous-workload-exec}
\end{figure}

Figure~\ref{continuous-workload-exec} shows the average execution times for Vanilla Linux/x86, Vanilla Linux/FPGA, and Xar-Trek for this periodic workload. (We again excluded  Vanilla Linux/ARM due to its inferior performance for the fixed workload case.) Xar-Trek outperforms Vanilla Linux/x86 and Vanilla Linux/FPGA by 18\% and 32\%, respectively.

Not surprisingly, Xar-Trek's gains in this case are relatively smaller than that in the fixed workload case for medium (Figure~\ref{Medium_Load}) and high (Figure~\ref{High_Load}) loads (where the largest gains were in the 88\%-31\% range) as those medium/high loads are not sustained here over time; they quickly decrease or increase. Nevertheless, the results show Xar-Trek scheduler's effectiveness for a time-varying load pattern. 

\begin{figure}[!ht]
\centering
\includegraphics[width=2.8in]{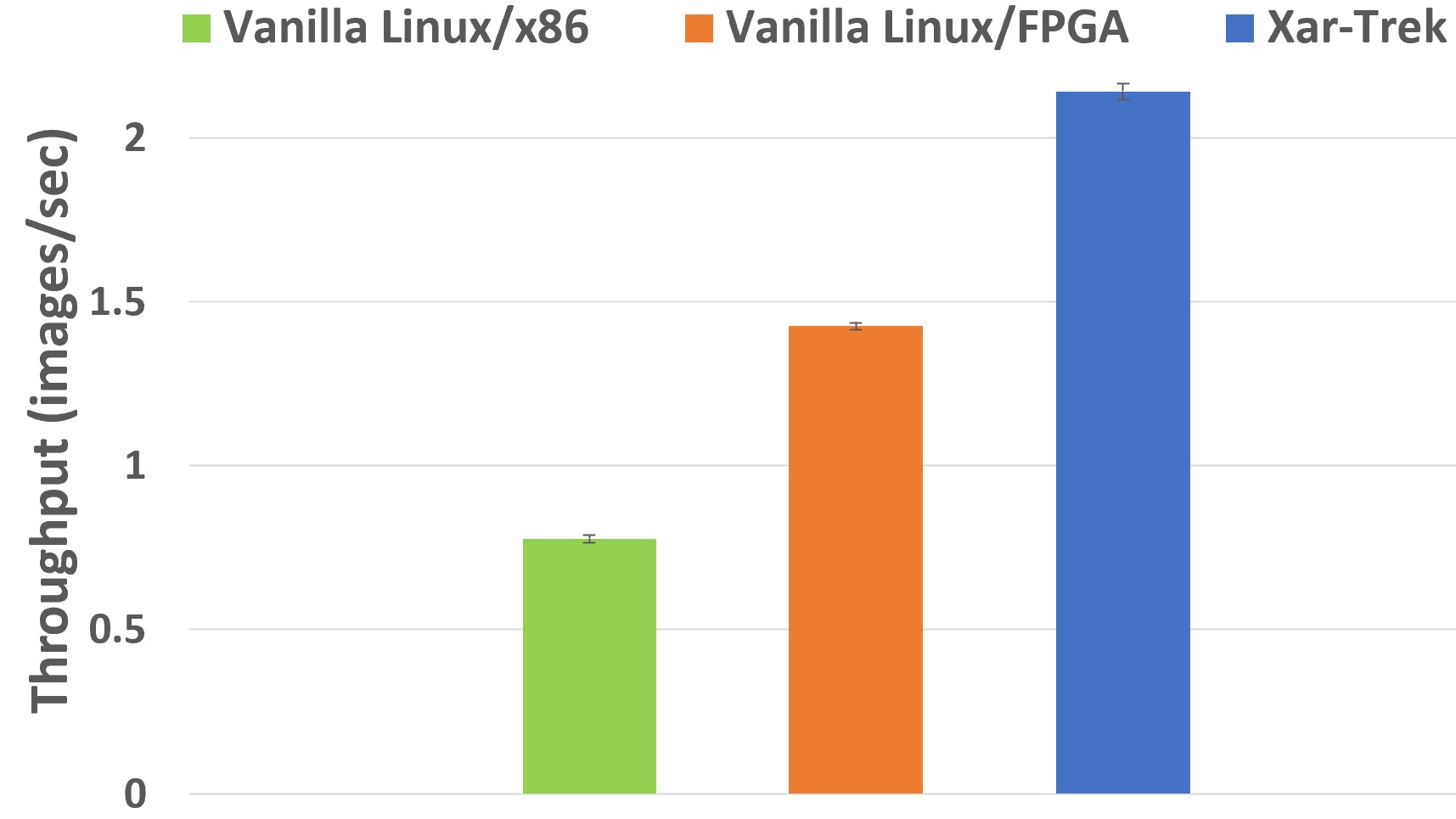}
\caption{Throughput of face detection under a periodic workload. Higher is better.} 
\label{continuous-workload-throughput}
\end{figure}

We conducted a similar periodic workload experiment to evaluate throughput. In this experiment, the periodic workload was generated by varying the number of processes from 10 (low) to 120 (high) during a time frame of 35 minutes. Figure~\ref{continuous-workload-throughput} shows the results. For each of the bars shown in the figure, the multi-face detection application was executed 10 times. Each run had a target of 1000 images to be processed and a duration of 60 seconds. At the end of each execution, the actual number of images processed were recorded and divided by 60 to obtain the throughput in images per second.

Figure~\ref{continuous-workload-throughput} shows that Xar-Trek outperforms Vanilla Linux/ FPGA and Vanilla Linux/x86 (50\% and 175\% respectively). The worst case, as expected, is for Vanilla Linux/x86. Again, compared to Figure~\ref{face-detect-throughput}, the throughput gains are relatively smaller here as the load changes over time. The results demonstrate Xar-Trek scheduler's effectiveness for obtaining superior throughput under time-varying loads.

\subsection{Profitable Workloads}

Not all applications can be profitably accelerated using FPGAs. Traditionally, the most profitable applications for FPGAs are those with compute-intensive operations such as floating point calculations, math operations, cryptographic computations, and matrix calculations, among others. (I/O-intensive applications are, of course, out of scope.) Applications with irregular memory access patterns generally yield inferior performance on FPGAs, especially on PCIe-attached FPGAs, largely because the FPGA can only operate on a limited set of local data at any given time~\cite{pointer-chasing-fpl16}.\footnote{This may not be always true on on-chip FPGAs   
with cache-coherent shared memory between CPUs and the FPGA (e.g.,~\cite{Xilinx_Versal,Intel_Agilex}).} Applications with pointer-chasing behaviors such as graph applications best exemplify this latter category. (In our five application set, CG\_A is the only one which exhibit this behavior to some degree; see Table~\ref{Exec_Time_App}.) This raises the question: for what percentage of compute-intensive applications in a given workload is Xar-Trek profitable?

\begin{table}[!ht]
\renewcommand{\arraystretch}{1.7}
\caption{Execution time of BFS application (milliseconds).}
\label{BFS-Exec_Time}
\centering
\begin{tabular}{c | c | c }
\hline
BFS's & x86 & FPGA \\
 number of nodes &  & \\
\hline
1,000 & 3.36  &  726.50  \\
\hline
2,000 & 115.74   & 2,282.54   \\
\hline
3,000 & 256.94   & 4,981.05   \\
\hline
4,000 & 458.04   & 8,760.80   \\
\hline
5,000 & 721.48  & 13,524.76 \\
\hline
\end{tabular}
\end{table}

To obtain insights to this question, we first  implemented a classical graph traversal algorithm, breadth-first-search (BFS), and measured its execution time on our x86 and FPGA hardware for graphs of different sizes (i.e., number of nodes). Table~\ref{BFS-Exec_Time} shows the results. For all graph sizes, x86 is faster by multiple orders of magnitude. (Our FPGA hardware, the  Xilinx Alveo U50 FPGA card, could not support graphs with larger than 5,000 nodes.) This means that, Xar-Trek's threshold estimation algorithm (Section~\ref{subsec:threshold-update}) will likely not find a reasonable CPU load that would justify migrating to the FPGA, and will almost always determine that the best target for the BFS function is x86. 

\begin{figure}[!ht]
\centering
\includegraphics[width=3.3in]{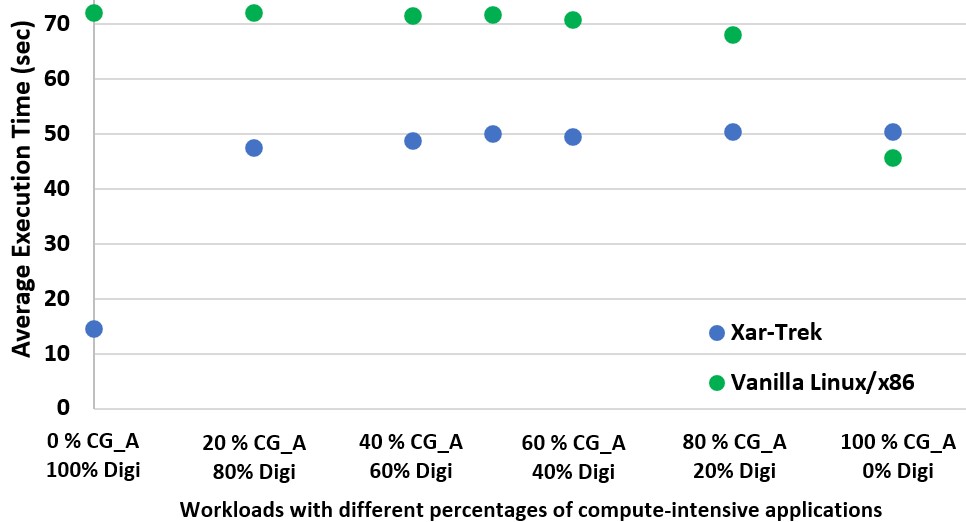}
\caption{Xar-Trek's effectiveness for different percentages of compute-intensive applications. Lower is better.} 
\label{sweet-spot}
\end{figure}

We now conducted an experiment where we fixed the load at 120 processes, varied the percentage of non-compute-intensive to compute-intensive applications from 0\% to 100\% in a ten-application set, and measured the average execution time under Xar-Trek and Vanilla Linux/x86. We used CG\_A as representative of a non-compute-intensive application since it is the slowest on the FPGA or the ARM CPU as Table~\ref{Exec_Time_App} shows. (We did not consider BFS for this experiment as Xar-Trek will always execute it on x86, as previously explained.) We used digit recognition (2000 tests) as representative of a compute-intensive application since it is one of the fastest on the FPGA (Table~\ref{Exec_Time_App}). We ran seven experiments with different percentages of CG\_A to digit recognition.

Figure~\ref{sweet-spot} shows the results. The figure's first workload-data point shows the case of only digit recognition applications, clearly illustrating Xar-Trek's  correct decision to run them on the FPGA. The last data point shows the case of only CG\_A applications, which favors Vanilla Linux/x86. The figure also shows that Xar-Trek always outperforms Vanilla Linux/x86 except for the last case (gains are in the 26\%-32\% range), revealing that Xar-Trek is profitable as long as the workload is dominated by compute-intensive applications. 

\subsection{Size of Binaries}

The development process of hardware accelerators includes the creation of two types of binary files: 1) executable and 2) hardware configuration (XCLBIN). The former is the binary generated by a traditional software compiler for CPUs and contains the implementation for a single-ISA CPU hardware, or is the binary generated by a multi-ISA compiler for heterogeneous-ISA CPU hardware~\cite{ASPLOS_2017_Breaking_Boundaries,ISCA_2014_ISA_Diversity_Chip,OSDI_2012_COMET}. The XCLBIN file contains the description of the hardware infrastructure that hosts an FPGA accelerator, the FPGA accelerator, and the interface with the host memory. 

Work in~\cite{ASPLOS_2017_Breaking_Boundaries,ISCA_2014_ISA_Diversity_Chip,OSDI_2012_COMET} shows that multi-ISA binary file sizes are larger than single-ISA file sizes since they contain code for each ISA. The traditional FPGA development process generates a binary file for a single-ISA host CPU (e.g., x86) and an XCLBIN file for the FPGA whose size depends on the size of the FPGA, available hardware kernels, etc. Since Xar-Trek's hardware includes multi-ISA CPUs and an FPGA, it generates both binary files.

\begin{figure}[!ht]
\centering
\includegraphics[width=3.0in]{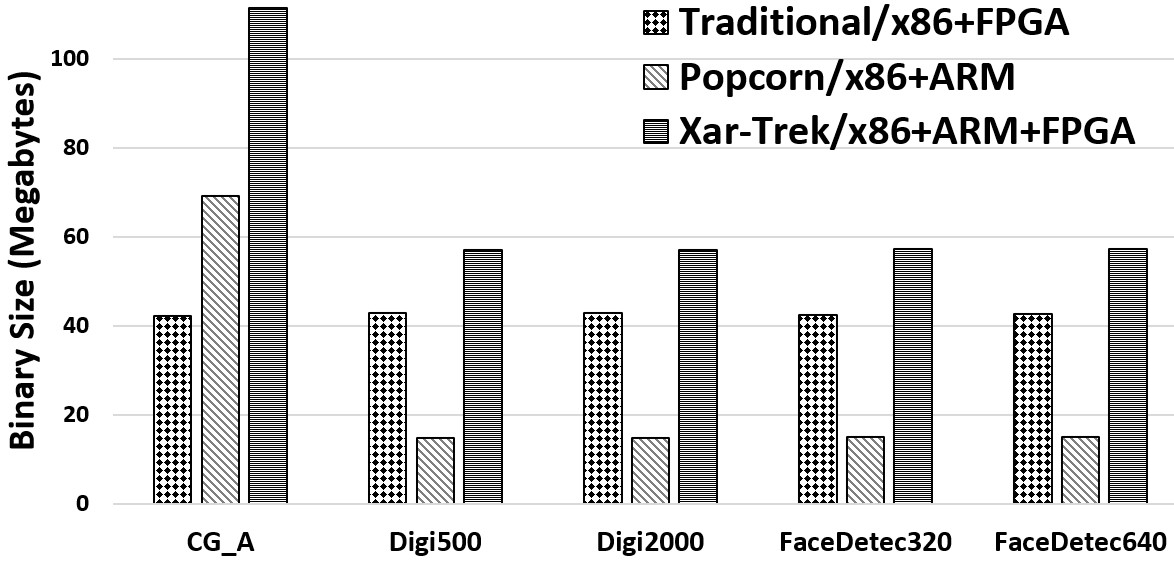}
\caption{Size of binaries. Smaller is better.} 
\label{binary-size}
\end{figure}

To understand the increase in the binary file sizes due to Xar-Trek's development process and the consequent storage and memory overheads, we compared Xar-Trek's file sizes against both baselines: i) the traditional FPGA development process (x86+FPGA) and ii) the heterogeneous-ISA development process using Popcorn Linux for x86+ARM hardware. 

Figure~\ref{binary-size} shows the total binary size (for the two files combined) for Xar-Trek and the two baselines for our five applications. Not surprisingly, Xar-Trek always has a larger binary size as it subsumes both the baselines (largest and smallest increases are 282\% and 33\%, respectively). 
Popcorn Linux has a larger size for the CG\_A application due to its 900 LOC, as opposed to 300-500 LOC for the face detection and digit recognition applications. Except for CG\_A, Xar-Trek does not impose a significant storage or memory overhead.

\section{Limitations}
\label{sec:limitations}


Xar-Trek has several limitations. First, execution migration to FPGAs is only possible for preselected application functions. This is largely a limitation of current state-of-the-art FPGA development tools and hardware, which restrict  hardware synthesis to self-contained functions with mostly CPU and memory operations (I/O operations are excluded). This limitation also reflects one of the motivations of using FPGAs: they are intended to mainly accelerate CPU-intensive code that operates on self-contained data, i.e., "compute kernels." This limitation also implies that execution migration is only possible at function boundaries -- a limitation that we acknowledged in Section~\ref{Compile-Time}. Functions can be outlined to bypass this limitation in some situations, e.g., when a function's logic is amenable to relatively easy decomposition. 

Another limitation that follows from the above is that execution migration is limited to application code; migrations inside shared libraries (e.g., glibc) and within OS system calls are not possible. This also reflects FPGAs' original motivation of accelerating compute kernels. Though there exists previous works on accelerating systems code (e.g., OS-kernel cryptography~\cite{DBLP:journals/corr/abs-1305-3345},  network-attached storage~\cite{10.1145/3315569}), it is unclear the degree to which this limitation reduces resource management flexibility in Xar-Trek's setting. 

Xar-Trek is limited to C. This limitation can be overcome and support for languages such as C++ can be added through additional engineering. State-of-the-practice FPGA development tools such as the Xilinx Vitis compiler do support C++. The Popcorn compiler used in Xar-Trek's multi-ISA binary generation step  (Section~\ref{Compile-Time}) also is limited to C. The major challenge in supporting languages such as C++ is generating multi-ISA code 
for advanced object-oriented features such as polymorphism and exception handling.


Xar-Trek is also limited to optimizing for performance. Due to that focus, we used server-grade, high-performance hardware including ThunderX ARM CPUs, which are not power-efficient. To additionally optimize for power, low-power embedded ARM boards can be used instead of server-grade ARM CPUs, such as what is done in~\cite{HDPC_19_HEXO}. Optimizing for power will also require computing metrics such as performance per watt~\cite{10.1145/957717.957772} or energy-delay-product~\cite{10.1109/40.888701,917539} to guide execution migration. Scheduling policies inspired by heuristics that balance power and performance in single-ISA heterogeneous settings (e.g.,~\cite{10.1145/2854038.2854047,179451}) can be designed.

\section{Related Work} \label{Related Work}

A large body of prior work has studied heterogeneous systems including CPU/FPGAs, CPU/GPUs, and heterogeneous-ISA CPUs. Our work most closely resembles prior work on application migration between heterogeneous-ISA CPUs, CPU/GPUs, and CPU/FPGAs. We compare and contrast Xar-Trek with the most closely related past works. 

The original body of work on heterogeneous-ISA CPUs~\cite{ASPLOS_2017_Breaking_Boundaries, EuroSys_15_Popcorn, ISCA_2014_ISA_Diversity_Chip, ASPLOS_12_ARM_MIPS_Migration,ACM_Trans_2015_K2,OSDI_2012_COMET} did not considers FPGAs. The latest in that space, HEXO \cite{HDPC_19_HEXO}, migrates  unikernel  virtual machines (VMs)~\cite{ASPLOS_13_UNIKERNEL} from x86 servers to embedded ARM boards to maximize throughput. HEXO uses a scheduler which decides when to migrate VMs based on an estimation of the slowdown the job would suffer if offloaded to the embedded system. HEXO also excludes FPGAs.

Lynx \cite{ASPLOS_20_Lynx_Smart_NIC} is an architecture that uses the CPU cores from an ASIC Smart NIC and the logic fabric from the FPGA smart NIC (presented in their previous work \cite{ATC_2019_NICA}) to offload server data and control planes from the host CPU. Lynx's performance improvement depends on the utilization of the eight ARM cores present on the ASIC smart NIC. In the case of an FPGA Smart NIC, Lynx's performance still depends on  the host CPU that is responsible for the application control. In contrast, Xar-Trek does not target hardware CPU cores inside an ASIC to execute the accelerated functions. These functions are implemented directly in the FPGA fabric, using available hardware acceleration tools to automatically transform candidate functions in hardware, not discarding the original functions that can run in software, if needed.

AIRA \cite{IEEE_Trans_Par_Dist_Syst_2018_AIRA} targets C/OpenMP applications and automatically instruments the source code to generate executables for CPUs and GPUs. The applications are analyzed offline in order to help an allocation policy to decide which architecture provides better performance. Subsequently, a partitioner creates specific ports to each architecture and a load balancer decides where to run the applications. 
AIRA  was evaluated on a CPU/GPU platform and demonstrated improvements over statically allocated applications. Xar-Trek follows AIRA's model of dynamic execution migration but focuses on application functions for execution migration. In addition, it allows the functions to be executed directly in hardware,  which AIRA excludes. 


Flick \cite{ISCA_20_FLICK} migrates application functions to an embedded processor inside an FPGA card connected to the host CPU through a PCIe interface. Targeting a shared memory approach, Flick uses a customized TLB and MMU (implemented on a microblaze processor), connected to a RISC-V core in the FPGA. A driver on the host is used to remap the TLB in the RISC-V core, with additional support from modifications to the Linux kernel, such as a page fault handler, kernel ELF loader, and a scheduler. After its execution on the processor core, the function returns to the host CPU. This execution model is similar to our work. However, in Xar-Trek, there is no need for customized embedded processor cores as the candidate functions are implemented directly in hardware. 

In \cite{FPL_16_Rec_Acc_Survey}, the authors present alternatives of hardware accelerators and compare them in terms of application, performance, energy, interface with the host CPU, design method, and integration. Applications are divided into two categories: batch, which process high volumes of data collected and stored in data centers, and streaming, which process high volumes of streaming data. The former benefits from improvements in throughput and the latter from improvements in latency. According to the authors, the vast majority of systems that utilize the FPGA as a co-processor accelerate functions that are only compute-intensive. Differently from the alternatives shown in \cite{FPL_16_Rec_Acc_Survey}, Xar-Trek allows the execution of the accelerated function also on the original CPU.

TornadoVM \cite{VEE_19_TornadoVM} is a virtualization layer implemented on top of Tornado \cite{VEE_17_Tornado}, a framework used for parallel programming. TornadoVM focuses on Java applications, evaluates their efficiency for different hardware platforms including CPUs, GPUs, and FPGAs at run-time, and makes optimal target selection decisions. 
In order to accomplish this, applications must be quickly compile-able for the available platforms. However, compilation times of FPGA development tools are in the order of minutes, compared to milliseconds for CPUs and GPUs. TornadoVM solves this problem by utilizing precompiled hardware modules. Although Xar-Trek is not focused on Java  applications, the hardware-accelerated functions are precompiled and available in the configuration bitstream, ready to be called by the scheduler when needed.

\section {Conclusions} \label {Conclusion}

Datacenters increasingly incorporate heterogeneous compute capabilities, from servers with GPUs to servers with FPGAs, to more recently, servers with ISA-diverse CPUs and FPGAs. An exemplar example of the last category is the emergence of (x86-based) servers equipped with "Smart" I/O devices such as SmartNICs and SmartSSDs that incorporate ARM- or RISC-V based CPUs as well as FPGAs. 

In datacenter's multi-tenant setting, server workloads can dynamically increase, degrading application performance. Usually, this  situation is alleviated by offloading work to other servers. Servers with heterogeneous-ISA CPUs and FPGAs provide a tantalizing alternative: dynamically migrate workloads from ISA-diverse CPUs to the FPGA. Our work shows that this is indeed possible and can yield compelling performance gains. The Xar-Trek framework's key capabilities that make this possible include a compiler, a run-time system, and a scheduling policy. The compiler generates multi-ISA binaries that incorporate FPGA implementations of select application functions; the run-time includes a scheduler mechanism that monitors server workloads and migrates  functions across ISA-diverse CPUs and FPGAs; and the scheduling policy decides what to migrate where. 

Cloud vendors have started offering FPGAs, starting 2016, notably with Amazon Web Services's F1 instances, using the traditional "pay-as-you-go" model: users wishing to accelerate their workloads with FPGAs pay for the acceleration capability and program the FPGAs with a hardware implementation of their workloads' compute kernels. Xar-Trek's vision leverages this investment in FPGAs, and exploits the FPGA -- when available -- to alleviate server workload spikes through transparent and dynamic execution migration, improving resource utilization and reducing ownership costs.

Several exciting directions for future work exist. One is to overcome Xar-Trek's limitations in Section~\ref{sec:limitations}.
Another is to space-share multiple applications concurrently on the FPGA to further improve Xar-Trek's resource utilization, possibly using the same circuit design to minimize context switching overheads, as in~\cite{8533480}. Yet another direction is developing defense mechanisms for Xar-Trek that increase security entropy against side channel exploits on the FPGA~\cite{szefer21}. Extending Xar-Trek's scheduler for the last two directions are interesting avenues for further research.

\begin{acks}
We thank the anonymous reviewers for their insightful comments which helped greatly improve the paper. We are grateful to the Xilinx University Program for their donation of the development tools used in this work. This work is supported by the US Office of Naval Research (ONR) under grants N00014-18-1-2022, N00014-19-1-2493, and N00014-21-1-2523.
\end{acks}

\bibliographystyle{ACM-Reference-Format}
\bibliography{ref}

\end{document}